# Ferromagnetism in transparent Mn(II)-doped indium tin oxide films prepared by sol-gel process

Susmita Kundu,[a)§] Dipten Bhattacharya,[b)] Jiten Ghosh,[c)] Pintu Das,[d)†] and Prasanta K. Biswas,[a)*]

[a)] *Sol-Gel Division, Central Glass and Ceramic Research Institute, Kolkata 700032, India*

[b)] *Sensor & Actuator Division, Central Glass and Ceramic Research Institute, Kolkata 700032, India*

[c)] *X-ray diffraction Facility, Central Glass and Ceramic Research Institute, Kolkata 700032, India*

[d)] *Institute of Experimental Physics, Saarland University, D-66041 Saarbrucken, Germany*

We observe remarkably strong room-temperature ferromagnetism [~1.5$\mu_B$/Mn(II)] in optically transparent Mn(II)-doped Indium Tin Oxide (ITO) films. The nanocrystalline films with average grain size 10-22 nm and thickness 150-350 nm are prepared by sol-gel coating technique on sodalime silica glass substrate. The ferromagnetic property is, of course, weak for films deposited on pure silica glass substrate. The structural parameters of the films appear to be governing the magnetic property strongly which vary appreciably depending on the substrate. The observation of room temperature ferromagnetism in transparent conducting ITO films may find a plethora of applications in the area of magneto-optics.

---

[*]Corresponding author; e-mail: pkbiswas@cgcri.res.in

[§]Currently at the Central Mechanical Engineering Research Institute, Durgapur, India

[†]Currently at the Max-Planck Institute for Chemical Physics of Solids, D-01187 Dresden, Germany

The subject of dilute magnetic semiconductor (DMS) has received a tremendous boost recently with the advent of spintronics-based applications where the spin degree of freedom of the electrons are utilized for performing almost the same jobs, performed by the semiconductor electronics, with even higher precision and coherence.[1] Observing room temperature ferromagnetism in DMS systems has always been the most important objective of all, for such applications. Recent successes[2-4] in the doping of magnetic ions within nano-crystals of different semiconductor systems resulted in nanocrystalline DMS with substantially improved optical and magnetic properties. Easy manipulation of spin via magnetic, electrical, and optical signals creates new possibilities of storing, processing, and transmitting information using spin-dependent currents. And the nanocrystalline DMS, certainly, offers hope for development of such spin-based devices with greater efficiency. While doping of different p- or n-type semiconductors – both non-oxide (II-VI or III-V) and oxide – by transition metal ions has been widely attempted so far,[5-11] which has yielded encouraging results including room temperature ferromagnetism, doping of optically active and nanocrystalline semiconductors for improved magnetic properties is still relatively less attempted.[12-13]

In this letter, we report that relatively low doping of Mn(II) ions (~6 at%) within nano-crystals (10-22 nm) of transparent indium tin oxide (ITO) films gives rise to a noticeable improvement in magnetic properties, e.g., room temperature ferromagnetism with negligible degradation in optical transparency due to optical absorption characteristics of Mn(II) ions in the visible range. Of course, such an improvement appears to depend crucially on the substrate of the films. While, films deposited on sodalime silica glass substrate exhibits ferromagnetism even at room temperature, the long-range ferromagnetic order weakens considerably in films coated on pure silica glass substrate. Such a variation in the magnetic property could result from variation in several structure parameters of the films – lattice parameters, lattice strain, Mn-O-Mn bond lengths, angles etc. – depending on the substrate.

The 6 at% Mn(II)-doped ITO films – $In_{1.69}Sn_{0.19}Mn_{0.12}O_3$ – have been deposited by sol-gel coating technique using a dip coater. The aquo-organic based precursor solution for the Mn(II) doped ITO films were prepared from hydrated indium and tin salts, $In(NO_3)_3.xH_2O$, $SnCl_4.5H_2O$, Mn acetate [$Mn(OAc)_2.4H_2O$], and Polyvinylalcohol (PVA) (molecular weight 22000, BDH, UK). The Mn salt and PVA were used as dopant and binder respectively. Initially an ITO precursor was prepared following the procedure as described in our previous report.[14] In the present case, Mn(II) doping was done by incorporation of requisite quantity of aqueous Mn(II) acetate into the ITO precursor maintaining the atomic ratio, In:Sn:Mn = 84.6:9.4:6.0. Next the solution was stirred for 2 h. Concentration of the sol - 6.0 wt% equivalent metal oxides -was maintained by adding water. The sol was aged for a few days before using for coating. The above precursor was used for the deposition of layers onto the cleaned bare sodalime silica glass and Heraeus (Germany) make suprasil grade pure silica glass following the dipping technique with a withdrawal speed 10 cm/min. The coated sample was placed in an air oven and heated to $100^0$C for 30 min. The dried film was then put into an electrical furnace and cured in air at a temperature $450^0 \pm 5^0$C for 30 min. The process of deposition follows a sequence of depositing 2-3 layers in order to improve the uniformity in the coverage and film thickness. The average film thickness varies within 150-350 nm according to the number of layers, 2 to 3. We report here the results of two films with 3 layers of coating prepared under identical conditions. The thickness of the films, in both the cases, is ~350 nm. We designate the samples depending on the types of substrates used for the deposition – sodalime silica glass [sample-A] and pure silica ($SiO_2$) glass [sample-B]. We observe a remarkable substrate-dependence of the magnetic properties, even though the Mn(II) doping level, curing temperature, and the film thickness are kept constant.

The films have been characterized by grazing incidence X-ray diffraction (GIXRD), atomic force microscopy (AFM), field emission scanning electron microscopy (FESEM) together with energy dispersive x-ray spectra (EDS), high resolution transmission electron microscopy (HRTEM), and surface characterization measurements such as surface roughness ($h_{cla}$), uniformity etc. In Fig. 1 we show the GIXRD spectra for both the samples A and B. The data were recorded in PANalytical X'Pert Pro MPD X-ray diffractometer in parallel beam geometry using thin film attachment with step size $0.05^0$ $(2\theta)$, step time 10s, and $2\theta = 10^0$-$80^0$. The microstructural parameters such as crystallite size, lattice strain, and lattice parameter were estimated for these samples from the Rietveld refinement of the XRD spectra by X'pert high score plus software (PANalytical) using the $Ia\bar{3}$ cubic space group symmetry (ICSD code:14387). The lattice parameters, lattice strain, and the reliability factors of the profile refinement are given in Table-I. The atomic positions of In, Sn, Mn, and O in the $In_2O_3$ lattice, as obtained from Rietveld refinement, are given in Table-II. The lattice parameter for the doped samples is smaller than that of pure $In_2O_3$ (a = 10.118 Å) as smaller Sn (0.83 Å) and Mn (0.80 Å) ions replace the In (0.94 Å) ions. For sample-A, both Sn and Mn dopant ions occupy the In2 (Wyckoff position: 8a) position in the lattice, where as for sample-B, Mn and Sn ions occupy the In1 (Wyckoff position: 24d) and In2 positions, respectively. Therefore, the order in which the Mn(II) ions are organized in the lattice does vary from sample to sample depending on the substrate. It is also noteworthy that while the $In_{1.8}Sn_{0.2}O_3$ films exhibit rhombohedral symmetry,[14] the Mn-doping tends to make the system cubic conforming to the structure of pure $In_2O_3$. The field emission scanning electron microscope (FESEM) and high-resolution transmission electron microscope (HRTEM) images (Fig. 1c) show that the films comprise of single crystalline nano-sized grains (10-22 nm) with presence of (222) planes. All these characterization results show clearly that the nano-sized crystals of the host $In_2O_3$ system have been successfully doped by both Sn and magnetic Mn(II) ions. This was recognized as a challenging task in the published literature.[4] The order of the distribution of dopant ions, of course, varies from sample to sample depending on the substrate.

We have carried out the study of optical spectra on all the films. The UV-VIS spectra of the films were recorded at room temperature using Shimadzu UV-VIS-NIR (model UV 3101 PC) spectrophotometer. Room temperature fluorescence (the lamp corrected for emission and excitation spectra of the sample) was recorded by Perkin-Elmer LS 55 Luminiscence spectrofluorometer. The UV-VIS spectra for both the samples A and B are shown in Fig. 2. Interestingly, nearly 80% transmittivity is retained even in the Mn(II)-doped films in visible range. Only a minor variation of the transmission spectra (Fig. 2b) could be observed. This is for the electronic transitions of Mn(II) – $^6A_1 \rightarrow {}^4E_G$ at ~485 nm [Refs. 14,15] and $^6A_1 \rightarrow {}^4T_{1G}$ at ~545 nm [Ref. 16].

The magnetization versus temperature patterns and hysteresis loops at different temperatures has been measured in a SQUID magnetometer (MPMS; Quantum Design) over a temperature range 10-300 K. In the main frame of Fig. 3a, we show the magnetization versus temperature patterns for sample-B while the hysteresis loops at low temperature are shown in the inset of Fig. 3a. In Fig. 3b, we show the hysteresis loops for sample-A. It is interesting to note that one observes a remarkable ferromagnetism even at room temperature in sample-A, where as for sample-B, the ferromagnetism is weakened considerably with no ferromagnetism at all at room temperature. We carried out Curie-Weiss fitting of the magnetization versus temperature pattern for sample-B: $\chi = C/(T-\theta)$ where $\chi$ = magnetic susceptibility = H/M (H = applied field, M = magnetization), C is the Curie constant and $\theta$ is the Weiss constant. The fitting is shown in the inset (i) of Fig. 3a. The Weiss constant $\theta$ is found to be positive (i.e., ferromagnetic) with a value of ~160 K. The Curie constant yields the value of magnetization ~0.08$\mu_B$/Mn(II) ions, which is quite small compared to the ferromagnetic moment per Mn(II) ion 5.92$\mu_B$ in $3d^5$ electronic state. Both these results show that the ferromagnetism in the sample-B is weak. Another point, worthy of mentioning here, is that there is hardly any divergence between zero-field (ZFC) and field cooled (FC) magnetization versus temperature plots. This shows that the magnetization over the entire temperature range (10-300 K) is not protocol or path dependent. The transition near ~160 K does not lead to a metastable phase. In the absence of high temperature magnetization data, such an analysis could not be carried out for sample-A. The hysteresis loops for sample-A depict saturation magnetization ($M_s$) as high as ~3.7$\mu_B$/Mn(II) ion at ~10 K which degrades down to ~1.5$\mu_B$/Mn(II) ion at 300 K. The room temperature hysteresis loop is shown in Fig. 3b inset. Interestingly, the ferromagnetism is retained right up to the room temperature. The coercivity varies within 75-250 Oe over the temperature range 10-300 K. The hysteresis loops are symmetric for both the samples and no signature of exchange-bias field could be noticed. Sizable coercivity over the entire temperature range 10-300 K shows that the system is not superparamagnetic.

The reason behind the ferromagnetism and its substrate dependence probably lies in the carrier mediated exchange interaction (RKKY) among the local spin structures of Mn(II) ions in this conducting oxides with high carrier concentration and mobility.[17] The detailed structure analysis shows that the average <Mn-O> bond length is ~1.9787 Å in sample-A where as it is ~2.1861 Å in sample-B. The RKKY exchange energy J depends significantly on the Mn-Mn separation. Therefore, J could be quite strong in the case of sample-A than in the case of sample-B yielding room temperature ferromagnetism in sample-A. Since, Sn and Mn ions occupy the In1 and In2 positions, respectively, in sample-B, there may be an ordered pattern in their distribution which could perhaps be absent for sample-A as both Sn and Mn ions occupy the In2 positions. Why and how the silica glass substrates give rise to such a different dopant ion distribution pattern – random and ordered – and hence different Mn(II)-Mn(II) bond length is yet to be fully understood. Further work is needed to unravel the substrate-film interface interaction.

In summary, we report room temperature ferromagnetism in highly transparent (~80%) 6 at% Mn(II)-doped ITO films deposited on sodalime silica glass. The Mn-O bond length and hence the Mn-Mn pair exchange interaction is found to depend significantly on the film substrate: smaller bond length and stronger exchange interaction in the case of films deposited on sodalime glass substrate and larger bond length and weaker exchange interaction in the case of films deposited on pure silica glass substrate. This work achieves both magnetic ion doping in nanocrystals of $In_2O_3$ as well as strong room temperature ferromagnetism without any degradation of optical transparency which could be useful for many opto-spintronics based applications.

This work is supported by a joint DST-JSPS program INT/P-5/JSPS/2006. We acknowledge experimental support of S. Banerjee and P. Mondal. We also acknowledge the help of Prof. A. Punnoose, Boise State University, USA for the HRTEM studies of the films.

[1]See, for example, S.A. Wolf, D.D. Awschalom, R.A. Buhrman, J.M. Daughton, S. von Molnàr, M.L. Roukes, A.Y. Chtchelkanova, and D.M. Treger, Science **294**, 1488 (2001); see also, I. Žutić, J. Fabian, and S. Das Sarma, Rev. Mod. Phys. **76**, 323 (2004).

[2]S.E. Erwin, L. Zu, L.I. Haftel, A.L. Efros, T.A. Kennedy, and D.J. Norris, Nature **436**, 91 (2005).

[3]D. Kitchen, A. Richardella, J-M. Tang, M.E. Flatté, and A. Yazdani, Nature **442**, 436 (2006).

[4]See, for example, A. Nag, S. Chakraborty, and D.D. Sarma, J. Am. Chem. Soc. (in press) and references therein.

[5]Y.-K. Zhou, M. Hashimoto, M. Kanamura, and H. Asahi, J. Supercond. **16**, 37 (2003).

[6]H. Saito, V. Zayets, S. Yamagata, and K. Ando, J. Appl. Phys. **93**, 6796 (2003).

[7]B. Vodungbo, Y. Zheng, F. Vidal, D. Demaille, V.H. Etgens, and D.H. Mosca, Appl. Phys. Lett. **90**, 062510 (2007).

[8]P. Sharma, A. Gupta, K.V. Rao, F.J. Owens, R. Sharma, R. Ahuja, J.M. Osorio, G.B. Johansson, and G.A. Gehring, Nature Mater. **2**, 673 (2003).

[9]D.P. Norton, S.J. Pearton, A.F. Hebard, N. Theodoropoulou, L.A. Boatner, and R.G. Wilson, Appl. Phys. Lett. **82**, 239 (2003).

[10]Y.G. Zhao *et al.*, Appl. Phys. Lett. **83**, 2199 (2003).

[11]Y. Matsumoto *et al.*, Science **291**, 840 (2001).

[12]S. B. Ogale, R. J. Choudhary, J. P. Buban, S. E. Lofland, S. R. Shinde, S. N. Kale, V. N. Kulkarni, J. Higgins, C. Lanci, J. R. Simpson, N. D. Browning, S. Das Sarma, H. D. Drew, R. L. Greene, and T. Venkatesan, Phys. Rev. Lett. **91**, 077205 (2003).

[13]N. Akdogan, H. Zabel, A. Nefedov, K. Westerholt, H-W Becker, S. Gök, R. Khaibullin, and L. Tagirov, e-print: arXiv.org/cond-mat/0807.4711 (2008).

[14]S. Kundu and P.K. Biswas, Chem. Phys. Lett. **432**, 508 (2006).

[15]K. Binghum and S. Parke, Phys. Chem. Glass **6**, 224 (1965).

[16]V. Ghioranescu, M. Sima, M.N. Grecu, and C. Ghica, Romanian Rep. Phys. **55**, 118 (2003).

[17]See, for example, M.J. Calderón and S. Das Sarma, e-print arXiv:cond-mat/0603182 (2006); see also, J.M.D. Coey, M. Venkatesan, and C.B. Fitzerald, Nature Mater. **4**, 173 (2005); J. Philip *et al*., Nature Mater. **5**, 298 (2006).

Table-I The values of lattice parameter, crystallite size, microstrain and reliability factors of the profile refinement as obtained from Rietveld Analysis.

| Sample ID | lattice parameter $a(A^0)$ | crystallite size (nm) | microstrain % | $R_p$ | $R_{wp}$ | $R_{exp}$ | GOF |
|---|---|---|---|---|---|---|---|
| Sample - A | 10.08107 | 17.81 | 0.72 | 28.83 | 38.50 | 35.43 | 1.180 |
| Sample – B | 10.10228 | 15.72 | 0.086 | 29.39 | 39.98 | 34.46 | 1.346 |

Table-II. The atomic coordinates of In1, In2, Mn, Sn, and O as obtained from the Rietveld refinement

| IO lattice where In1 was replaced by Mn | | | | IO lattice where In2 was replaced by Mn | | | | IO lattice where In1 was replaced by Sn | | | | IO lattice where In2 was replaced by Sn | | | |
|---|---|---|---|---|---|---|---|---|---|---|---|---|---|---|---|
| Wyckoff Position | Atomic position | | | Wyckoff Position | Atomic position | | | Wyckoff Position | Atomic position | | | Wyckoff Position | Atomic position | | |
| | x | y | z | | x | y | z | | x | y | z | | x | y | z |
| **Sample – A** | | | | | | | | | | | | | | | |
| Mn 24 d | 0.2963 | 0.0000 | 0.2500 | In 24 d | 0.2857 | 0.0000 | 0.2500 | Sn dopant ions do not occupy the In1 | | | | In 24 d | 0.2807 | 0.0000 | 0.2500 |
| In 8 a | 0.0000 | 0.0000 | 0.0000 | Mn 8 a | 0.0000 | 0.0000 | 0.0000 | Position in IO lattice (shown in | | | | Sn 8 a | 0.0000 | 0.0000 | 0.0000 |
| O 48 e | 0.0977 | 0.3591 | 0.0000 | O 48 e | 0.0977 | 0.0890 | 0.1347 | Fig.1) | | | | O 48 e | 0.0977 | 0.3591 | 0.1347 |
| **Sample – B** | | | | | | | | | | | | | | | |
| Mn 24 d | 0.3540 | 0.0000 | 0.2500 | Mn dopant ions do not occupy the In2 | | | | Sn 24 d | 0.2837 | 0.0000 | 0.2500 | In 24 d | 0.2770 | 0.0000 | 0.2500 |
| In 8 a | 0.0000 | 0.0000 | 0.0000 | Position in IO lattice (shown in | | | | In 8 a | 0.0000 | 0.0000 | 0.2500 | Sn 8 a | 0.0000 | 0.0000 | 0.0000 |
| O 48 e | 0.5839 | 0.3591 | 0.2488 | Fig.1) | | | | O 48 e | 0.0977 | 0.3591 | 0.1347 | O 48 e | 0.1200 | 0.3591 | 0.1347 |

## Figure Captions

Fig. 1. (color online) The x-ray diffraction patterns as well as their Rietveld refinement for the Mn-doped ITO films – sample A (a) and sample B (b); (c) the FESEM Image of the Sample-A; inset: the grain growth pattern could be seen; inset shows the HRTEM image of lattice fringes of (222) plane and the histogram of grain size distribution.

Fig. 2. (color online) (a) The optical absorption spectra for the sample-A and sample-B; (b) the transmissivity in visible range for sample-A and an undoped ITO sample.

Fig. 3. (color online) (a) The dc magnetization versus temperature pattern – both ZFC and FC – for the sample-B; the black symbols depict the data corresponding to the left axis while the blue symbols depict the data corresponding to the right axis; insets: (i) the Curie-Weiss plot and (ii) the hysteresis loops at different temperatures; (b) the hysteresis loop for sample-A at 10 K; inset: the hysteresis loop for the sample-A at 300 K.

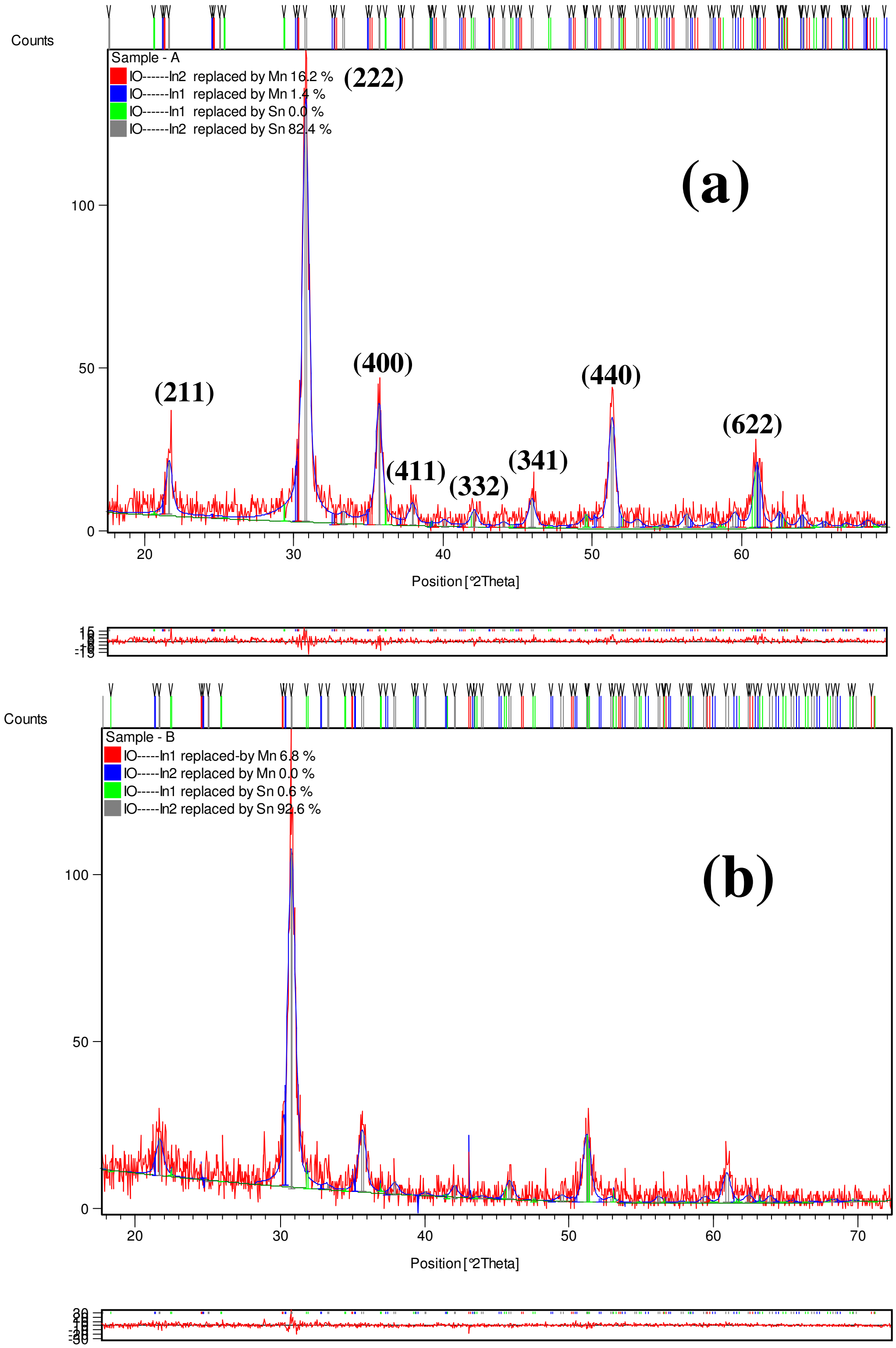

**Fig. 1. S. Kundu *et al.***

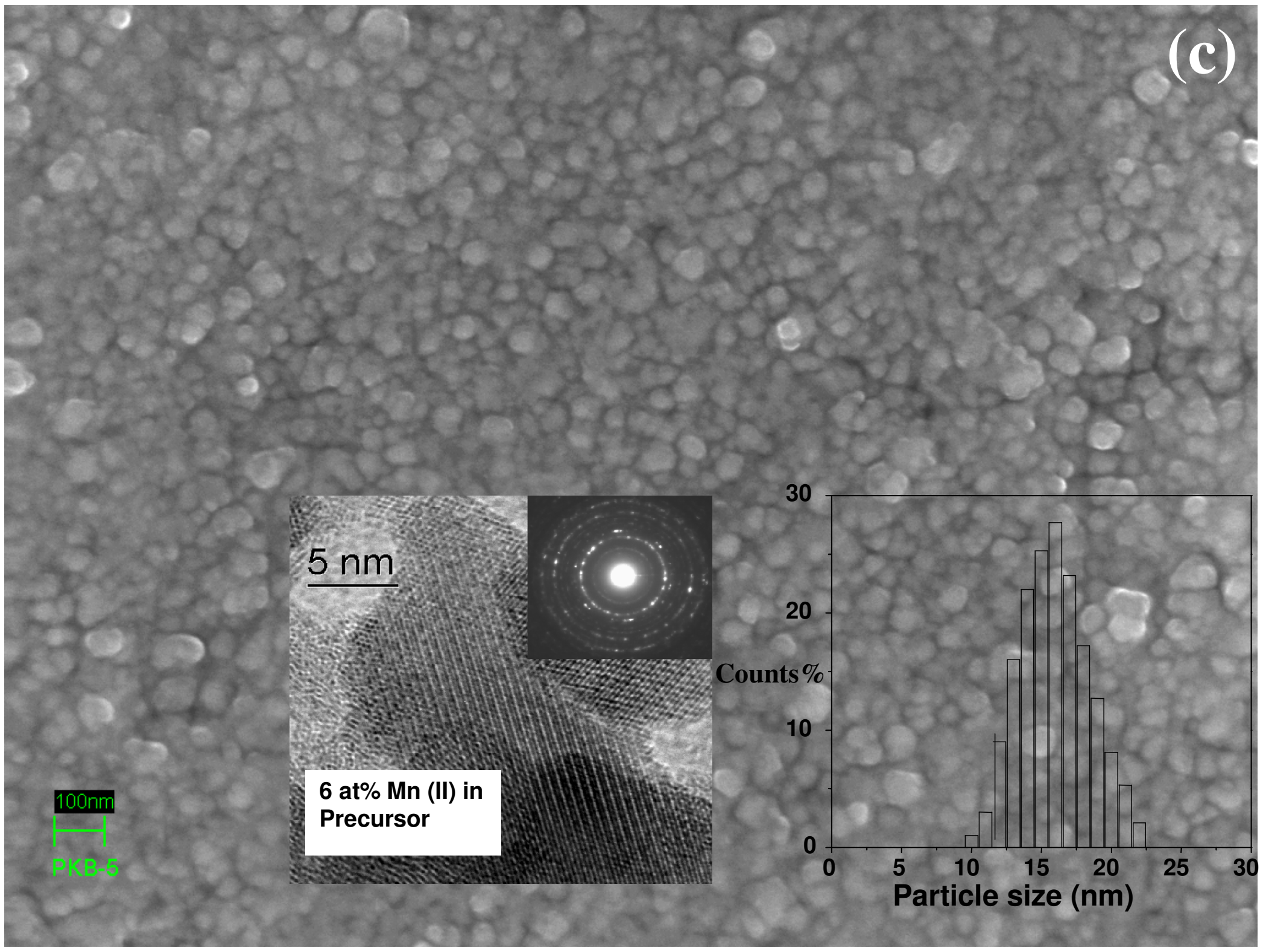

**Fig. 1. S. Kundu *et al*.**

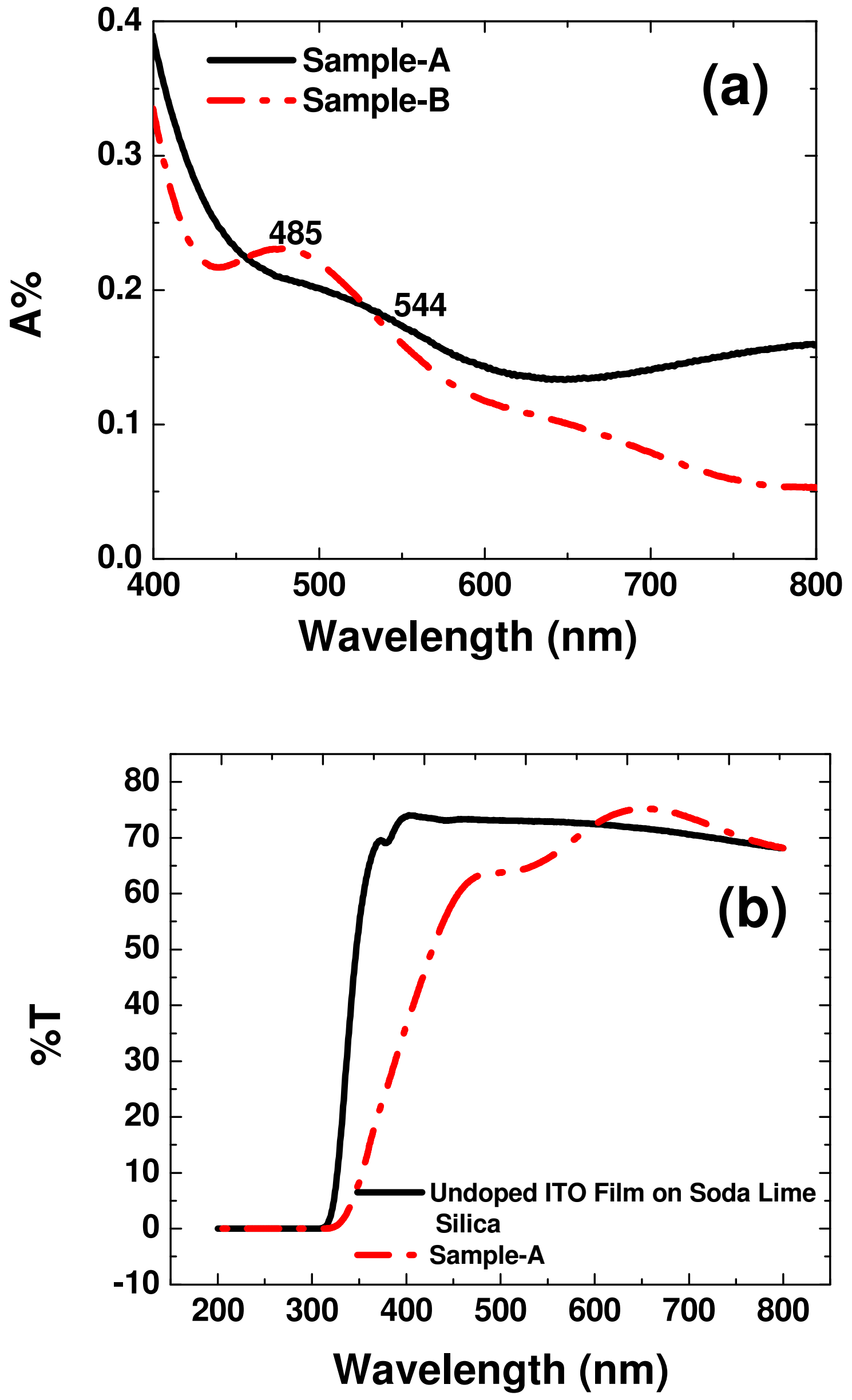

**Fig. 2. S. Kundu *et al*.**

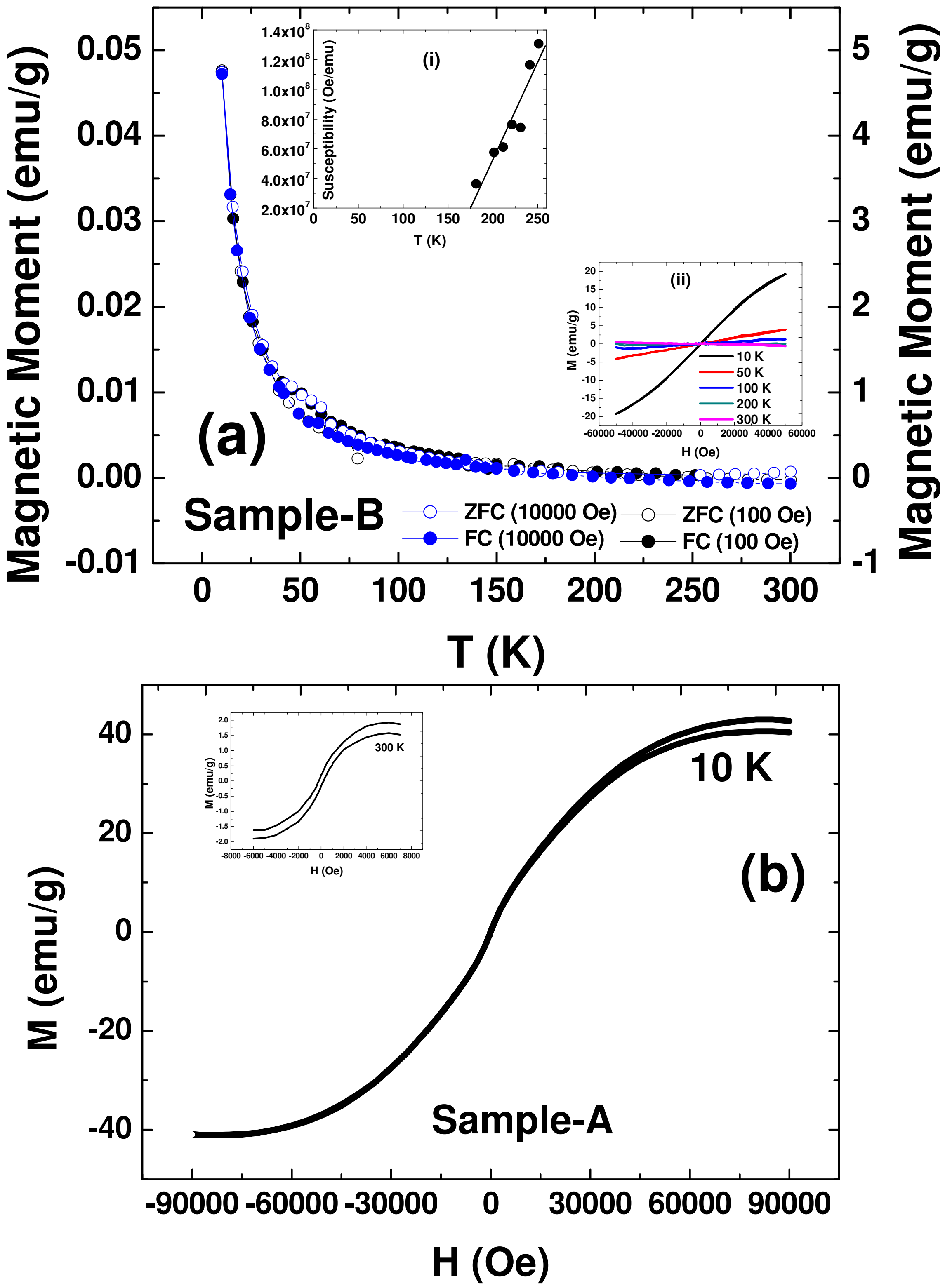

Fig. 3. S. Kundu *et al*.